\documentclass[journal,twocolumn,10pt]{IEEEtran}
\usepackage{graphicx}
\usepackage{float}
\usepackage{amsfonts}
\usepackage{amsmath}
\usepackage{amssymb}
\usepackage{cite}
\usepackage{bm}
\usepackage{mathrsfs}
\newtheorem{proposition}{Proposition}
\newtheorem{corollary}{Corollary}

\newpage

\begin{document}
\title{Large System Decentralized Detection Performance Under
Communication Constraints}

\author{Sudharman K. Jayaweera,~\IEEEmembership{Member,~IEEE}\thanks{The author is with
the Department of Electrical and Computer Engineering, Wichita
State University, Wichita, KS, 67260, USA. E-mail:
sudharman.jayaweera@wichita.edu.}
\thanks{This research was supported by Kansas National Science
Foundation EPSCOR program under the grants KUCR
\#NSF32223/KAN32224 and KUCR \#NSF32241. This work was presented
at the IEEE Vehicular Technology Conference, Stockholm, Sweden,
May 2005.} }

\markboth{To appear in IEEE Communications Letters, S. K.
Jayaweera, June, 2005}{To appear in IEEE Communications Letters,
S. K. Jayaweera, June, 2005}

\maketitle

\begin{abstract}
The problem of decentralized detection in a sensor network
subjected to a total average power constraint and all nodes
sharing a common bandwidth is investigated. The bandwidth
constraint is taken into account by assuming non-orthogonal
communication between sensors and the data fusion center via
direct-sequence code-division multiple-access (DS-CDMA). In the
case of large sensor systems and random spreading, the asymptotic
decentralized detection performance is derived assuming
independent and identically distributed (iid) sensor observations
via random matrix theory. The results show that, even under both
power and bandwidth constraints, it is better to combine many
not-so-good local decisions rather than relying on one (or a few)
very-good local decisions.
\end{abstract}


\section{Introduction}
This paper considers decentralized detection in
energy-constrained, large wireless sensor networks in noisy,
band-limited channels. Although there is a considerable amount of
previous work on the subject of distributed detection, most of it
used to ignore the effect of noisy channels between the local
sensors and data fusion center. Even less is the attention
received by bandlimited noisy channels in the context of
decentralized detection. For example, while distributed detection
performance of an energy-constrained wireless sensor network over
a noisy channel has been considered recently \cite{chamberland3},
it assumes orthogonal sensor-to-fusion center communication
leading to an infinite bandwidth assumption. However, in
applications involving dense, low-power, distributed wireless
sensor networks it is more likely that all nodes will share a
common available bandwidth. In this case, the assumption of large
sensor systems implies non-orthogonal communication between the
sensor nodes and the fusion sensor.

An important design objective in low-power wireless sensor systems
is to extend the whole network lifetime. Thus, a sensible
constraint on the sensor system is a finite total power
\cite{chamberland3}. In this paper, the bandwidth constraint is
taken into account by assuming non-orthogonal direct-sequence
code-division multiple-access (DS-CDMA) communication between
sensors and the data fusion center. The main contribution of this
paper is the derivation of the decentralized detection
performance, in closed-form, under a total power constraint when
the communication channel between the local sensors and the fusion
center is both bandlimited and noisy. As we will see, the
performance is a function of the exact signalling codes used by
the distributed sensors for any finite-size sensor network.
However, in the case of random spreading we are able to derive an
elegant and simple closed-form expression that is independent of
the exact spreading codes once we consider asymptotically large
sensor systems. This is our main result and, as we will see, it
allows us to draw general conclusions regarding the design of
wireless sensor systems under such total power constraints in
noisy and bandlimited channels.

The remainder of the paper is organized as follows: In Section
\ref{sec:system_model} we present our system model. Next, in
Section \ref{sec:analysis} we use random matrix theory to derive a
closed-form expression for the decentralized detection performance
in a large sensor system followed by a discussion of our analysis.
Finally, in Section \ref{sec:conclusions} we conclude by
summarizing our results.

\section{System Model Description} \label{sec:system_model}

We consider a binary hypothesis testing problem in an $N_s$-node
wireless sensor network connected to a data fusion center via
distributed parallel architecture. Let us denote by $H_{0}$ and
$H_{1}$ the null and alternative hypotheses, respectively, having
corresponding prior probabilities $P(H_0) = p_{0}$ and $P(H_1) =
p_{1}$. We will consider that the observed stochastic process
under each hypothesis consists of one of two possible Gaussian
signals, denoted by $X_{0,n}$ and $X_{1,n}$, corrupted by additive
white Gaussian noise. Under the two hypotheses the $n$-th local
sensor observation $z_n$, for $n = 1,\cdots N_s$, can be written
as
\begin{eqnarray}
H_{0} : \ \ \ \ z_{n} &=& X_{0,n} + v_n \nonumber \\
H_{1} : \ \ \ \ z_{n} &=& X_{1,n} + v_n  \label{eq:local01}
\end{eqnarray}
where the observation noise $v_n$ is assumed to be zero-mean
Gaussian with the collection of noise samples having a covariance
matrix $\Sigma_v$. Each local sensor processes its observation
$z_{n}$ independently to generate a local decision $u_{n}(z_n)$
which are sent to the fusion center. Let us denote by ${\bf
r}(u_{1}(z_1), u_{2}(z_2), \cdots, u_{N_s}(z_{N_s}))$ the received
signal at the fusion center. The fusion center makes a final
decision based on the decision rule $u_0({\bf r})$. The problem at
hand is to choose $u_0({\bf r}), u_{1}(z_1), u_{2}(z_2), \cdots,
u_{N_s}(z_{N_s})$ so that a chosen performance metric is
optimized.

The solution to this problem is known to be too complicated under
the most general conditions \cite{tenney1}. While optimal local
processing schemes have been derived under certain special
assumptions, a class of especially important local processers are
those that simply amplify the observations before retransmission
to the fusion center \cite{chamberland3,chamberland2}. Thus, the
local sensor decisions sent to the fusion center are given by,
$u_{n} = g z_n$ for $n = 1,\cdots N_s$ where $g>0$ is the analog
relay amplifier gain at each node. In our model all sensor nodes
share a common bandwidth and a total available energy. For
analytical reasons, as well as due to their practical relation to
DS-CDMA communications, we consider bandwidth sharing
non-orthogonal communication based on spreading in which each
sensor node is assigned a signature code of length $N$. If the
$n$-th sensor node is assigned the code ${\bf s}_n$, the received
chip-matched filtered and sampled discrete-time signal at the
fusion center can be written as ${\bf r} = g \sum_{n=1}^{N_s} {\bf
s_n} z_n + {\bf w} = g {\bf S} {\bf z} + {\bf w}$ where ${\bf r}$
and ${\bf w}$ are $N$-dimensional received signal and receiver
noise vectors, respectively and the $n$-th column of the $N \times
N_s$ matrix ${\bf S}$ is equal to the vector ${\bf s}_n$. We
assume that the receiver noise is a white Gaussian noise process
so that the filtered noise vector ${\bf w} \sim {\cal N} ({\bf 0},
\sigma_w^2 {\bf I}_N)$.Then we have that
\begin{eqnarray}
H_{0} : \ \ \ \ {\bf r} &\sim& {\cal N} \left({\bf m}_0, \Sigma_0
\right) \nonumber \\
H_{1} : \ \ \ \ {\bf r} &\sim& {\cal N} \left({\bf m}_1, \Sigma_1
\right) \label{eq:local06}
\end{eqnarray}
where, for $j = 0, 1$, ${\bf m}_j = g {\bf S} {\mathbb E} \{ {\bf
X}_j \}$ and $\Sigma_j = g^2 {\bf S} \left( Cov ({\bf X}_j ) +
\Sigma_v \right) {\bf S}^T + \sigma_w^2 {\bf I}_N$.

To be specific, consider the detection of a deterministic signal
so that ${\bf X}_1 = - {\bf X}_0 = m {\bf 1}$ is known ($m > 0$)
and $\Sigma_0 = \Sigma_1 = \Sigma$ where (${\bf 1}$ is the vector
of all ones) $\Sigma = g^2 {\bf S} \Sigma_v {\bf S}^T + \sigma_w^2
{\bf I}_N$. With these assumptions we also have that ${\bf m}_1 =
- {\bf m}_0 \ = \ g m {\bf S} {\bf 1}$ and the radiated power of
node $n$ is then given by ${\mathbb E} \{ |u_{n}|^2 \} = g^2
{\mathbb E} \{ |z_{n}|^2 \} = g^2 (m^2 + \sigma_v^2)$ where
$\sigma_v^2$ is the observation noise variance. Let us define the
total power constraint the whole sensor system is subjected to as
$P$, so that the amplifier gain $g$ is given by
\begin{eqnarray}
g &=& \sqrt{\frac{P}{N_s (m^2 + \sigma_v^2)}} . \label{eq:local04}
\end{eqnarray}
Then, it can be shown that the optimal threshold rule at the
fusion center is of the form
\begin{eqnarray}
u_{0}(\textbf{r}) &=& \left\{\begin{array}{ccccc} 1   & & &\geq& \\
&\textrm{if}& T \left( \textbf{r} \right) & & \tau' \\
0  & & &<&
\end{array}\right. , \label{eq:local10}
\end{eqnarray}
where we have defined the decision variable $T$ as $T \left(
\textbf{r} \right) = \left( {\bf m}_1 - {\bf m}_0 \right)^T
\Sigma^{-1} \textbf{r} = 2 g m {\bf 1}^T {\bf S}^T \left(  g^2
{\bf S} \Sigma_v {\bf S}^T + \sigma_w^2 {\bf I}_N \right)^{-1}
\textbf{r}$ and $\tau'$ is the threshold that depends on the
specific optimality criteria. It can be shown that the false-alarm
$P_f$ and miss $P_m$ probabilities of the detector
(\ref{eq:local10}) are given by
\begin{eqnarray}
P_f &=& Q \left( \frac{\tau' + 2 g^2 m^2 {\bf 1}^T {\bf S}^T
\Sigma^{-1} {\bf S} {\bf 1}}{2 g m \sqrt{{\bf 1}^T {\bf S}^T
\Sigma^{-1} {\bf S} {\bf 1}}} \right) , \label{eq:local11}
\end{eqnarray}
and
\begin{eqnarray}
P_m &=& Q \left( \frac{2 g^2 m^2 {\bf 1}^T {\bf S}^T \Sigma^{-1}
{\bf S} {\bf 1} - \tau'}{2 g m \sqrt{{\bf 1}^T {\bf S}^T
\Sigma^{-1} {\bf S} {\bf 1}}} \right) . \label{eq:local12}
\end{eqnarray}
For example, in the case of Neyman-Pearson optimality at the
fusion center, $\tau'$ is chosen to minimize $P_m$ subject to an
upper bound on $P_f$. On the other hand under Bayesian minimum
probability of error optimality one would choose $\tau'$ to
minimize $P_e = p_0 P_f + p_1 P_m$. As one would expect, the
performance of course depends on the particular codes assigned to
each sensor node as seen from (\ref{eq:local11}) and
(\ref{eq:local12}). Thus, while it is possible to evaluate the
performance for specific systems via (\ref{eq:local11}) and
(\ref{eq:local12}), it is rather difficult to draw general
conclusions regarding the design of decentralized detection
systems. However, such conclusions can be reached for large
systems through asymptotic analysis, as we show next.

\section{Large Sensor System Performance Analysis} \label{sec:analysis}

Let us assume that the spreading codes are chosen randomly so that
each element of ${\bf s}_n$ takes either $\frac{1}{\sqrt{N}}$ or
$-\frac{1}{\sqrt{N}}$ with equal probability. Moreover, we take
independent sensor observations such that $\Sigma_v = \sigma_v^2
{\bf I}$. Let us assume a large sensor system such that both $N_s$
and $N$ are large such that $\lim_{N \longrightarrow \infty}
\frac{N_s}{N} = \alpha$. Now using a theorem on the convergence of
the empirical distribution of eigenvalues of a large random matrix
proven in \cite{evans1}, we may prove the following proposition,
which is the main result of this paper:

\proposition{With ${\bf S}$ and $\Sigma$ defined as above,
\begin{eqnarray}
g^2 {\bf 1}^T {\bf S}^T \Sigma^{-1} {\bf S} {\bf 1}
&\longrightarrow& \left( \frac{\sigma_v^2}{N_s} + \frac{m^2 +
\sigma_v^2}{P \beta_0} \right)^{-1} , \label{eq:local17}
\end{eqnarray}
almost surely, as $N \longrightarrow \infty$, where
\begin{small}
\begin{eqnarray}
\beta_0 = \frac{\sqrt{ \left( \gamma + \sigma_w^2 \right)^2
\alpha^2 + 2 \gamma (\sigma_w^2 - \gamma) \alpha + \gamma^2} -
\left(\gamma + \sigma_w^2 \right) \alpha + \gamma}{2 \gamma
\sigma_w^2} \label{eq:local18}
\end{eqnarray}
\end{small}
with $\gamma = \frac{P}{N} \left(1+ \frac{m^2}{\sigma_v^2}
\right){-1}$ and $\Sigma_v = \sigma_v^2 {\bf I}$.}
\label{prop:prop01}

\proof{See Appendix I.}

The proposition \ref{prop:prop01} leads to the following corollary
on the asymptotically large sensor system performance of
decentralized detection in noisy bandlimited channels:

\corollary{With all notation as defined above, when $\lim_{N
\longrightarrow \infty} \frac{N_s}{N} = \alpha$, the large sensor
network performance of the decentralized detection is given by
$P_f \longrightarrow  Q \left(\sqrt{\mu}(\tau' + \frac{2
m^2}{\mu})/2 m \right)$ and $P_m \longrightarrow  Q
\left(\sqrt{\mu} (\frac{2 m^2}{\mu} - \tau')/2 m \right)$ where
$\mu = \frac{\sigma_v^2}{N_s} + \frac{m^2 + \sigma_v^2}{P
\beta_0}$.
\begin{figure*}[htbp!] \centerline{\hbox{
\begin{tabular}{cc}
\includegraphics[height=2in]{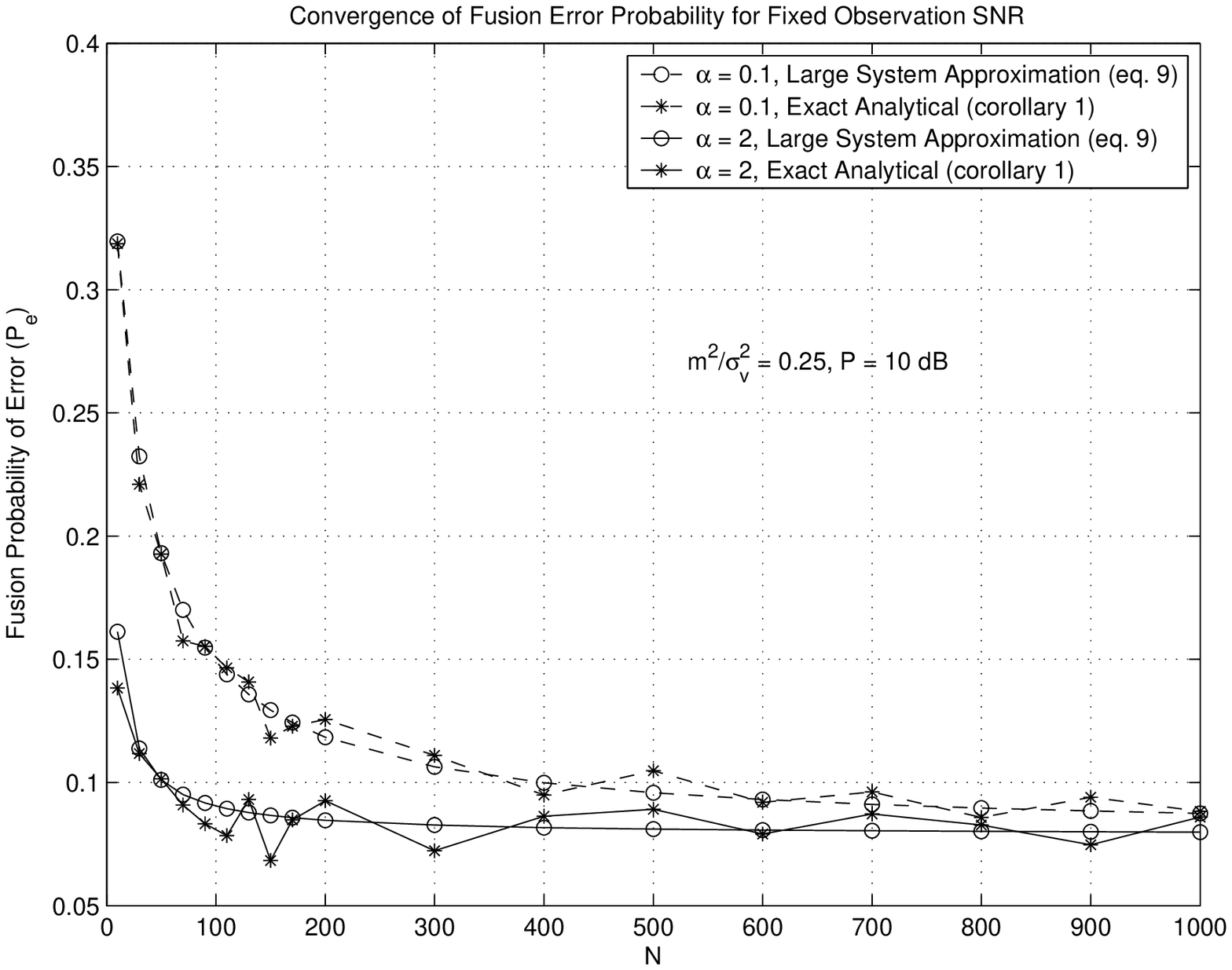} &
\includegraphics[height=2in]{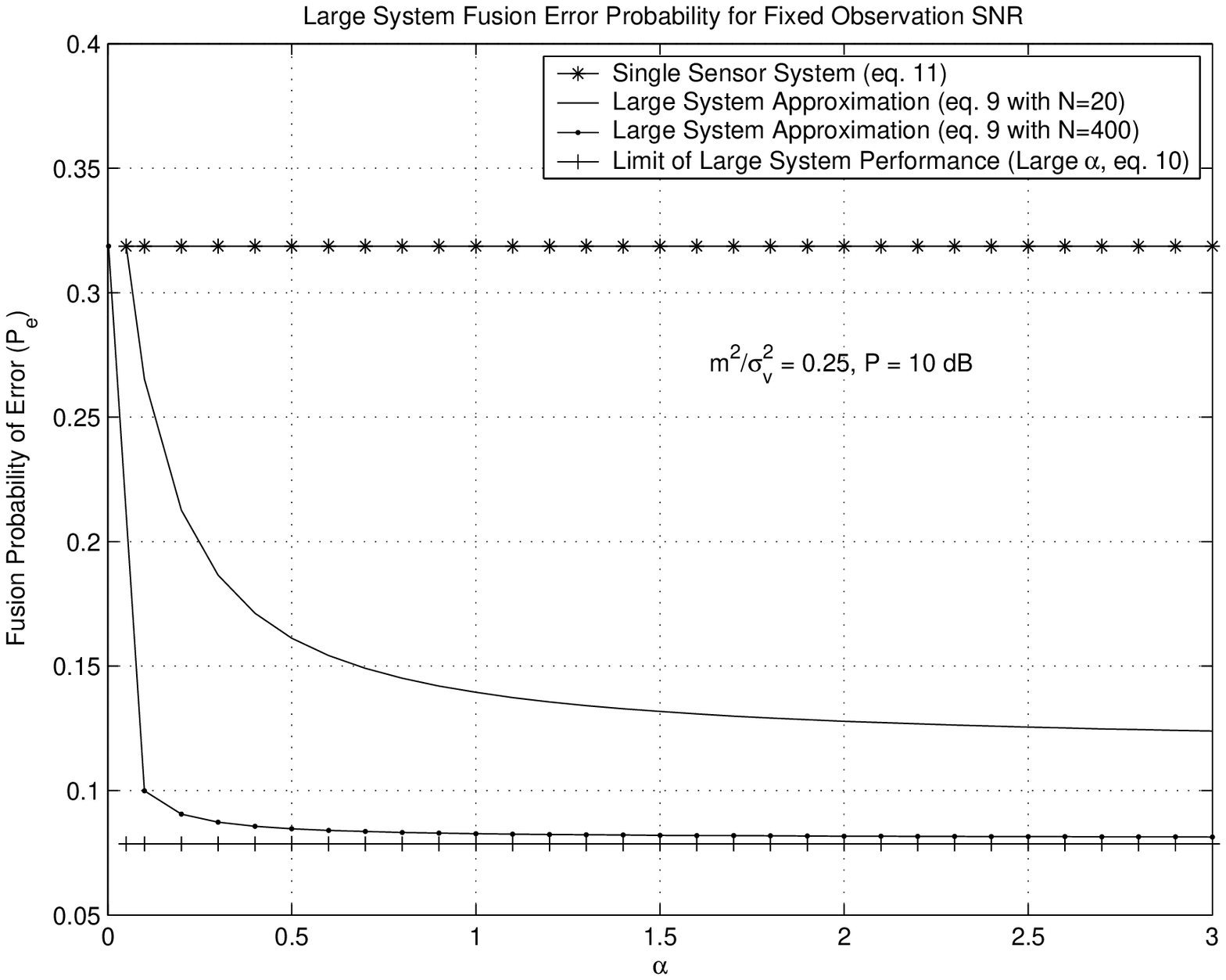} \\
(a) & (b) \end{tabular} }} \caption{Decentralized Detection
Performance in a Noisy, Bandlimited Channel Subjected to a Total
Power Constraint (a) Large Sensor System Approximation (b) Limit
of Large Sensor System Approximation when $\alpha \longrightarrow
\infty$.} \label{fig:fig01}
\end{figure*}

The above corollary leads to insights on large sensor system
performance of decentralized detection in noisy, bandlimited
channels. For instance, in the special case of minimum probability
of error optimality at the fusion center, according to corollary
1, the large system probability of error is asymptotically given
by
\begin{eqnarray}
P_e (\alpha) &\longrightarrow& Q \left( m/\sqrt{\mu} \right) ,
\label{eq:local19}
\end{eqnarray}
where convergence is almost surely and $\mu$ is as defined above.

Figure \ref{fig:fig01}a shows the convergence of the
random-spreading based decentralized detection performance as
predicted by (\ref{eq:local19}). Note that the exact analysis in
Fig. \ref{fig:fig01}a was obtained for a random choice of the code
matrix ${\bf S}$. As can be seen from Fig. \ref{fig:fig01}a,
(\ref{eq:local19}) provides a good approximation to the detection
performance for large spreading lengths $N$, and thus for
large-sensor systems (since $N_s=N \alpha$). More importantly, we
can observe from Fig. \ref{fig:fig01}a that for each fixed $N$,
increasing $\alpha$ improves the decentralized detection
performance. Since this is equivalent to increasing the number of
sensors $N_s$ allowed in the system for a fixed bandwidth we
conclude that it is better to allow as many sensors to send their
local decisions to the fusion center.

In fact, for large alpha, one can show that $\beta_0
\longrightarrow \frac{1}{\sigma_w^2}$, and as a result, in this
case the error probability in (\ref{eq:local19}) goes to
\begin{eqnarray}
P_e (\alpha)  &\longrightarrow&  Q \left( \sqrt{
\frac{P}{\sigma_w^2} \left(1+\frac{\sigma_v^2}{m^2}\right)^{-1}}
\right) . \label{eq:local20}
\end{eqnarray}
On the other hand, if one were to allocate all available power $P$
and the total bandwidth to just one sensor node the fusion center
performance will be given by
\begin{eqnarray}
P_{e,1}  &=&  Q \left( \sqrt{\frac{P}{\sigma_w^2} \left(
\frac{P/\sigma_w^2}{m^2/ \sigma_v^2} + \left(1+
\frac{\sigma_v^2}{m^2}\right) \right)^{-1}} \right) .
\label{eq:local21}
\end{eqnarray}

Comparison of (\ref{eq:local20}) and (\ref{eq:local21}) shows that
allowing more sensor nodes in the network is better even if the
channel is both noisy and bandlimited. This comparison is shown in
Fig. \ref{fig:fig01}b. First, we can observe from Fig.
\ref{fig:fig01}b that as $N$ increases the fusion center
performance improves. Secondly we see that as $N \longrightarrow
\infty$, the performance for large $\alpha$ indeed goes to
(\ref{eq:local20}). Third, Fig. \ref{fig:fig01}b confirms that
combining more local decisions is better than allocating all
available power and bandwidth to one sensor. Moreover, the
performance improves monotonically with increasing $\alpha$ (for a
fixed $N$) showing that it is better to combine as many local
decisions as possible at the fusion center. We should divide the
available power among all nodes and allow all of them to share the
available bandwidth even if they are to interfere with each other
due to non-orthogonality.

\section{Conclusions} \label{sec:conclusions}
We analyzed the decentralized detection performance of a total
average power constrained wireless sensor network in a noisy and
bandlimited channel. Assuming that the sensors-to-fusion center
communication is based on DS-CDMA, a closed form expression for
the fusion performance, and its large system asymptotic under
random spreading were derived. It was shown that in a noisy,
bandlimited channel it is beneficial to combine as many sensor
local decisions as possible even if this leads to non-orthogonal
sensor-to-fusion center communication.

\appendices
\section{The Proof of Proposition \ref{prop:prop01}}

\proof{Using the definitions of ${\bf S}$ and ${\bf 1}$, we can
write
\begin{small}
\begin{eqnarray}
g^2 {\bf 1}^T {\bf S}^T \Sigma^{-1} {\bf S} {\bf 1} = g^2 \left(
\sum_{n=1}^{N_s} {\bf s}_n^T \Sigma^{-1} {\bf s}_n +
\sum_{n=1}^{N_s} \sum_{\underset{n' \neq n}{n'=1}}^{N_s} {\bf
s}_n^T \Sigma^{-1} {\bf s}_{n'} \right) \label{eq:proof11}
\end{eqnarray}
\end{small}
Let ${\cal I}$ denote a set of sensor indices (i.e. ${\cal I}
\subset \{ 1, 2, \cdots, N_s \}$), ${\bf S}_{\cal A}$ denote the
matrix ${\bf S}$ with column indices specified by set ${\cal A}$
deleted, ${\bf \Lambda}_n = g^2 \sigma_v^2 {\bf I}_n$ and ${\bf
Q}_{\cal A} = \left( {\bf S}_{\cal A} {\bf \Lambda}_{{N_s}-|{\cal
A|}} {\bf S}_{\cal A} + \sigma_w^2 {\bf I}_N \right)$ where ${\bf
I}_n$ and $|{\cal A}|$ are the $n \times n$ identity matrix and
the cardinality of set ${\cal A}$, respectively. Then, for $n=1,
\cdots, N_s$, using the matrix inversion lemma we can show that
${\bf s}_n^T \Sigma^{-1} {\bf s}_n = {\bf s}_n^T {\bf
Q}_{\{n\}}^{-1} {\bf s}_n/(1+g^2 \sigma_v^2 {\bf s}_n^T {\bf
Q}_{\{n\}}^{-1} {\bf s}_n)$. But, applying Theorem 7 of
\cite{evans1} and using (\ref{eq:local04}), we can show that ${\bf
s}_n^T {\bf Q}_{\{n\}}^{-1} {\bf s}_n \longrightarrow \beta_0$
almost surely, where $\beta_0$ is as given by (\ref{eq:local18})
and $\gamma = \frac{P}{N} \left(1+ \frac{m^2}{\sigma_v^2}
\right)^{-1}$. Combining these we have almost surely
\begin{eqnarray}
{\bf s}_n^T \Sigma^{-1} {\bf s}_n &\longrightarrow&
\left({\beta_0}^{-1} + g^2 \sigma_v^2 \right)^{-1} .
\label{eq:proof14}
\end{eqnarray}
Similarly, repeated application of matrix inversion lemma twice
show that,
\begin{small}
\begin{eqnarray}
{\bf s}_n^T \Sigma^{-1} {\bf s}_{n'} = \frac{{\bf s}_n^T {\bf
Q}_{\{n, n'\}}^{-1} {\bf s}_{n'}}{\left(1+g^2 \sigma_v^2 {\bf
s}_n^T {\bf Q}_{\{n\}}^{-1} {\bf s}_{n}\right) \left(1+g^2
\sigma_v^2 {\bf s}_{n'}^T {\bf Q}_{\{n, n'\}}^{-1} {\bf
s}_{n'}\right)} . \label{eq:proof15}
\end{eqnarray}
\end{small}
Now the use of Theorem 7 of \cite{evans1} shows that RHS goes to
zero almost surely, for $n\neq n'$. Substituting
(\ref{eq:proof14}) and (\ref{eq:proof15}) in (\ref{eq:proof11})
gives (\ref{eq:local17}), completing the proof. }

\bibliographystyle{IEEEtran}
\bibliography{Distributeddetection}
\end{document}